\documentclass[preprint,12pt]{elsarticle}

\newcommand{\Rey}{\text{Re}}
\newcommand{\Pen}{\text{Pe}}

\newtheorem{problem}{Problem}

\usepackage{listings}
\usepackage{color}

\definecolor{dkgreen}{rgb}{0,0.6,0}
\definecolor{gray}{rgb}{0.5,0.5,0.5}
\definecolor{mauve}{rgb}{0.58,0,0.82}

\usepackage{tikz}
\usetikzlibrary{backgrounds}
\usepackage{graphicx}
\usepackage{bm}
\usepackage{amsmath}
\usepackage{setspace}
\usepackage{hyperref}
\usepackage{subfigure}
\usepackage{multirow}
\hypersetup{
    colorlinks=true,       
}
\usepackage{amssymb}

\usepackage{lineno}
\usepackage{diagbox}

\biboptions{sort&compress}

\journal{Journal Name}

\begin{document}

\begin{frontmatter}


\title{Deep Learning of Vortex Induced Vibrations}



\author{Maziar Raissi$^{1}$, Zhicheng Wang$^{2}$, Michael S. Triantafyllou$^{2}$,\\ and George Em Karniadakis$^{1}$}

\address{$^{1}$Division of Applied Mathematics, Brown University, Providence, RI, 02912, USA\\
$^{2}$Department of Mechanical Engineering, Massachusetts Institute of Technology,\\ Cambridge, MA 02139, USA}

\begin{abstract}
Vortex induced vibrations of bluff bodies occur when the vortex shedding frequency is close to the natural frequency of the structure. Of interest is the prediction of the lift and drag forces on the structure given some limited and scattered information on the velocity field. This is an inverse problem that is not straightforward to solve using standard computational fluid dynamics (CFD) methods, especially since no information is provided for the pressure. An even greater challenge is to infer the lift and drag forces given some dye or smoke visualizations of the flow field. Here we employ deep neural networks that are extended to encode the incompressible Navier-Stokes equations coupled with the structure's dynamic motion equation. In the first case, given scattered data in space-time on the velocity field and the structure's motion, we use four coupled deep neural networks to infer very accurately the structural parameters, the entire time-dependent pressure field (with no prior training data), and reconstruct the velocity vector field and the structure's dynamic motion. In the second case, given scattered data in space-time on a concentration field only, we use five coupled deep neural networks to infer very accurately the vector velocity field and all other quantities of interest as before. This new paradigm of inference in fluid mechanics for coupled multi-physics problems enables velocity and pressure quantification from flow snapshots in small subdomains and can be exploited for flow control applications and also for system identification.
\end{abstract}


\end{frontmatter}



\section{Introduction}

Fluid-structure interactions (FSI) are omnipresent in engineering applications \cite{MP_book_1998,MP_book_2004}, e.g. in long pipes carrying fluids, in heat exchangers, in wind turbines, in gas turbines, in oil platforms and long risers for deep sea drilling. Vortex induced vibrations (VIV), in particular, are a special class of fluid-structure interactions (FSI), which involve a resonance condition. They are caused in external flows past bluff bodies when the frequency of the shed vortices from the body is close to  the natural frequency of the structure \cite{Williamson_2004}. 
A prototypical example is flow past a circular cylinder that involves the so-called von K\'{a}rm\'{a}n shedding with a non-dimensional frequency (Strouhal number) of about $0.2$. If the cylinder is elastically mounted, its resulting motion is caused by the lift force and the drag force in the crossflow and streamwise directions, respectively, and can reach about $1D$ and $0.1D$ in amplitude, where $D$ is the cylinder diameter. Clearly, for large structures like a long riser in deep sea drilling, this is a very large periodic motion that will lead to fatigue and hence a short life time for the structure.\\

Using traditional computational fluid dynamics (CFD) methods we can predict accurately VIV (\cite{Evangelinos_VIV}), both the flow field and the structure's motion.  However, CFD simulations are limited to relatively low Reynolds numbers and simple geometric configurations and involve the generation of elaborate and moving grids that may need to be updated frequently. Moreover, some of the structural characteristics, e.g. damping, are not readily available and hence separate experiments are required to obtain such quantities involved in CFD modeling of FSI. Solving inverse coupled CFD problems, however, is typically computationally prohibitive and often requires the solution of ill-posed problems. For the vibrating cylinder problem, in particular, we may have available data for the motion of the cylinder or some limited noisy measurements of the velocity field in the wake or some flow visualizations obtained by injecting dye upstream for liquid flows or smoke for air flows. Of interest is to determine the forces on the body that will determine the dynamic motion and possibly deformation of the body, and ultimately its fatigue life for safety evaluations.\\

In this work, we take a different approach building on our previous work on physics informed deep learning \cite{raissi2017physics_I,raissi2017physics_II} and extending this concept to coupled multi-physics problems. Instead of solving the fluid mechanics equations and the dynamic equation for the motion of the structure using numerical discretization, we learn the velocity and pressure fields and the structure's motion using coupled deep neural networks with scattered data in space-time as input. The governing equations are employed as part of the loss function and play the role of regularization mechanisms. Hence, experimental input that may be noisy and at scattered spatio-temporal locations can be readily utilized in this new framework. Moreover, as we have shown in previous work, {\em physics informed neural networks} are particularly effective in solving inverse and data assimilation problems in addition to leveraging and discovering the hidden physics of {\em coupled multi-physics} problems \cite{raissi2018deep}.\\

This work aims to demonstrate feasibility and accuracy of a new approach that involves  Navier-Stokes informed deep neural networks inference, and is part of our ongoing development of {\em physics-informed learning machines} \cite{raissi2017physics_I,raissi2017physics_II}. The first glimpses of promise for exploiting structured prior information to construct data-efficient and physics-informed learning machines have already been showcased in the recent studies of \cite{raissi2017inferring, raissi2017machine, owhadi2015bayesian}. There, the authors employed Gaussian process regression \cite{Rasmussen06gaussianprocesses} to devise functional representations that are tailored to a given linear operator, and were able to accurately infer solutions and provide uncertainty estimates for several prototype problems in mathematical physics. Extensions to nonlinear problems were proposed in subsequent studies by Raissi {\em et. al.} \cite{raissi2018numerical, raissi2018hidden} in the context of both inference and systems identification. Despite the flexibility and mathematical elegance of Gaussian processes in encoding prior information, the treatment of nonlinear problems introduces two important limitations. First, in \cite{raissi2018numerical,raissi2018hidden} a local linearization in time of the nonlinear terms in time is involved, thus limiting the applicability of the proposed methods to discrete-time domains and possibly compromising the accuracy of their predictions in strongly nonlinear regimes. Secondly, the Bayesian nature of Gaussian process regression requires certain prior assumptions that may limit the representation capacity of the model and give rise to robustness/brittleness issues, especially for nonlinear problems  \cite{owhadi2015brittleness}, although this may be overcome by hybrid neural networks/Gaussian process algorithms \cite{pang2018neural}.\\

Physics informed deep learning \cite{raissi2017physics_I,raissi2017physics_II,raissi2018deep} takes a different approach by employing deep neural networks and leveraging their well known capability as universal function approximators \cite{hornik1989multilayer}. In this setting, one can directly tackle nonlinear problems without the need for committing to any prior assumptions, linearization, or local time-stepping. Physics informed neural networks exploit recent developments in automatic differentiation \cite{baydin2015automatic} -- one of the most useful but perhaps under-utilized techniques in scientific computing -- to differentiate neural networks with respect to their input coordinates and model parameters to leverage the underlying physics of the problem. Such neural networks are constrained to respect any symmetries, invariances, or conservation principles originating from the physical laws that govern the observed data, as modeled by general time-dependent and nonlinear partial differential equations. This simple yet powerful construction allows us to tackle a wide range of problems in computational science and introduces a potentially transformative technology leading to the development of new data-efficient and physics-informed learning machines, new classes of numerical solvers for partial differential equations, as well as new data-driven approaches for model inversion and systems identification.\\

Here, we should underline an important distinction between this line of work and existing approaches (see e.g., \cite{beidokhti2009solving}) in the literature that elaborate on the use of machine learning in computational physics. The term {\em physics-informed machine learning} has been also recently used by Wang {\it et. al.} \cite{wang2017comprehensive} in the context of turbulence modeling. Other examples of machine learning approaches for predictive modeling of physical systems include \cite{zhu2018bayesian,hagge2017solving,tripathy2018deep,vlachas2018data,parish2016paradigm,duraisamy2015new,ling2016reynolds,zhang2015machine,milano2002neural,perdikaris2016multifidelity,rico1994continuous,ling2015evaluation}. All these approaches employ machine learning algorithms like support vector machines, random forests, Gaussian processes, and feed-forward/convolutional/recurrent neural networks merely as {\em black-box} tools. As described above, our goal here is to open the black-box, understand the mechanisms inside it, and utilize them to develop new methods and tools that could potentially lead to new machine learning models, novel regularization procedures, and efficient and robust inference techniques. To this end, the proposed work draws inspiration from the early contributions of Shekari Beidokhti and Malek \cite{beidokhti2009solving}, Psichogios and Ungar \cite{psichogios1992hybrid}, Lagaris {\em et. al.} \cite{lagaris1998artificial}, as well as the contemporary works of Kondor \cite{kondor2018n,kondor2018generalization}, Hirn \cite{hirn2017wavelet}, and Mallat \cite{mallat2016understanding}.\\

The focus of this paper is fluid-structure interactions (FSI) in general and vortex induced vibrations (VIV) in particular. Specifically, we are interested in the prediction of the fluid's lift and drag forces on the structure given some limited and scattered information on the velocity field or simply snapshots of dye visualization. This is a data assimilation problem that is notoriously difficult to solve using standard computational fluid dynamics (CFD) methods, especially since no initial conditions are specified for the algorithm, the training domain is small leading to the well-known numerical artifacts, and no information is provided for the pressure. An even greater challenge is to infer the lift and drag forces given some dye or smoke visualizations of the flow field. Here we employ Navier-Stokes informed deep neural networks that are extended to encode the structure's dynamic motion equation. The paper is organized as follows. In the next section, we give an overview of the proposed algorithm, set up the problem and describe the synthetic data that we generate to test the performance of the method. In section \ref{sec:Results}, we present our results for three benchmark cases; (1) We start with a pedagogical example and assume that we know the forces acting on the body while seeking to obtain the structure's motion in addition to its properties without explicitly solving the equation for displacement. (2) We then consider a case where we know the velocity field and the structural motion at some scattered data points in space-time. In this case, we try to infer the lift and drag forces while learning the pressure field as well as the entire velocity field in addition to the structure's dynamic motion. (3) We then consider an even more interesting case where we only assume the availability of concentration data in space-time. From such information we obtain all components of the flow fields and structure's motion as well as lift and drag forces. We conclude with a short summary.

\section{Problem Setup and Solution Methodology}

We begin by considering the prototype VIV problem of flow past a circular cylinder. The fluid motion is governed by the incompressible Navier-Stokes equations while the dynamics of the structure is described in a general form involving displacement, velocity, and acceleration terms. In particular, let us consider the two-dimensional version of flow over a flexible cable, i.e., an elastically mounted cylinder \citep{Bourguet_2011_JFM}. The two-dimensional problem contains most of the salient features of the three-dimensional case and consequently it is relatively straightforward to generalize the proposed framework to the flexible cylinder/cable problem. In two dimensions, the physical model of the cable reduces to a mass-spring-damper system. There are two directions of motion for the cylinder: the streamwise (i.e., $x$) direction and the crossflow (i.e., $y$) direction. In this work, we assume that the cylinder can only move in the crossflow (i.e., $y$) direction; we concentrate on crossflow vibrations since this is the primary VIV direction. However, it is a simple extension to study cases where the cylinder is free to move in both streamwise and crossflow directions.

\subsection{A Pedagogical Example}

The cylinder displacement is defined by the variable $\eta$ corresponding to the crossflow motion. The equation of motion for the cylinder is then given by
\begin{equation}\label{eq:Structure}
\rho \eta_{tt} + b \eta_t + k \eta = f_L,
\end{equation}
where $\rho$, $b$, and $k$ are the mass, damping, and stiffness parameters, respectively. The fluid lift force on the structure is denoted by $f_L$. The mass $\rho$ of the cylinder is usually a known quantity; however, the damping $b$ and the stiffness $k$ parameters are often unknown in practice. In the current work, we put forth a deep learning approach for estimating these parameters from measurements. We start by assuming that we have access to the input-output data $\{t^n, \eta^n\}_{n=1}^N$ and $\{t^n, f_L^n\}_{n=1}^N$ on the displacement $\eta(t)$ and the lift force $f_L(t)$ functions, respectively. Having access to direct measurements of the forces exerted by the fluid on the structure is obviously a strong assumption. However, we start with this simpler but pedagogical case and we will relax this assumption later in this section.\\

\begin{figure}[!t]
\centering
\includegraphics[width=0.75\textwidth]{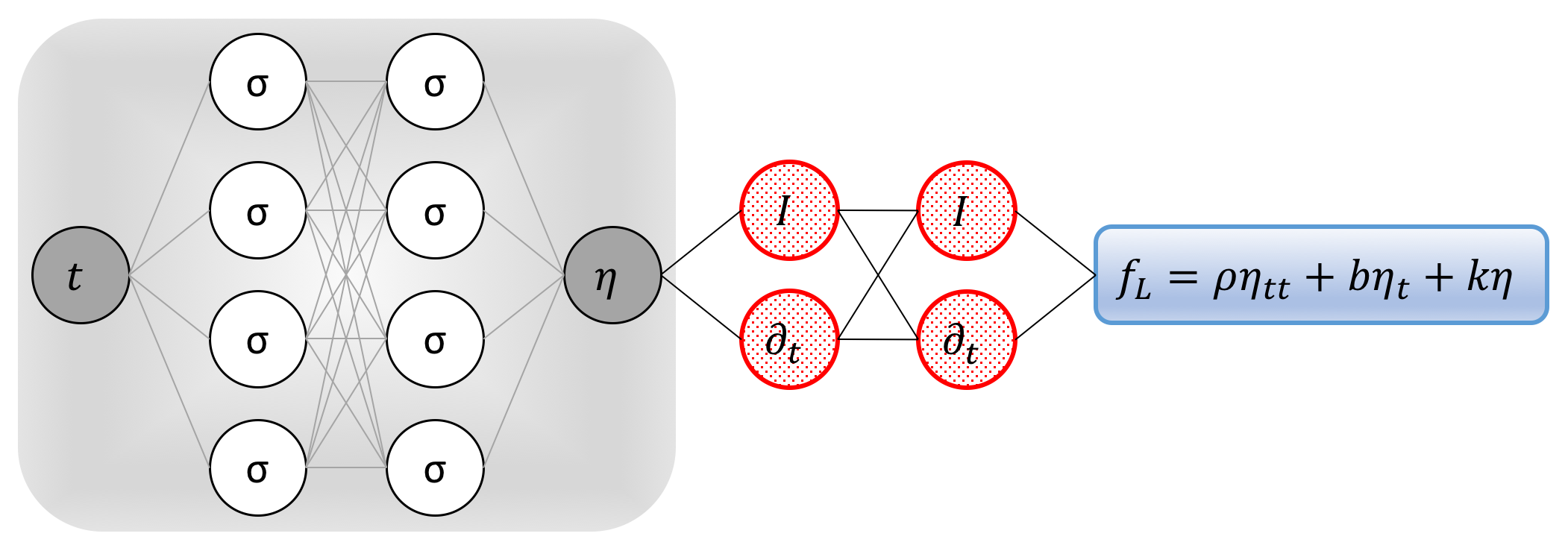}
\caption{\emph{Pedagogical physics-informed neural network:} A plain vanilla densely connected (physics uninformed) neural network, with 10 hidden layers and 32 neurons per hidden layer per output variable (i.e., $1 \times 32 = 32$ neurons per hidden layer), takes the input variable $t$ and outputs $\eta$. As for the activation functions, we use $\sigma(x) = \sin(x)$. For illustration purposes only, the network depicted in this figure comprises of 2 hidden layers and 4 neurons per hidden layers. We employ automatic differentiation to obtain the required derivatives to compute the residual (physics informed) networks $f_L$. The total loss function is composed of the regression loss of the displacement $\eta$ on the training data, and the loss imposed by the differential equation $f_L$. Here, $I$ denotes the identity operator and the differential operator $\partial_t$ is computed using automatic differentiation and can be thought of as an ``activation operator". Moreover, the gradients of the loss function are back-propagated through the entire network to train the parameters of the neural network as well as the damping $b$ and the stiffness $k$ parameters using the Adam optimizer.}\label{fig:DeepVIV_1}
\end{figure}

Inspired by recent developments in \emph{physics informed deep learning} \citep{raissi2017physics_I,raissi2017physics_II} and \emph{deep hidden physics models} \citep{raissi2018deep}, we propose to approximate the unknown function $\eta$ by a deep neural network. This choice is motivated by modern techniques for solving forward and inverse problems involving partial differential equations, where the unknown solution is approximated either by a neural network \citep{raissi2017physics_I,raissi2017physics_II,raissi2018deep,raissi2018forwardbackward,raissi2018multistep} or a Gaussian process \citep{raissi2018numerical,raissi2018hidden,raissi2017inferring,raissi2017machine,raissi2017parametric,perdikaris2017nonlinear,raissi2016deep}. Moreover, placing a prior on the solution is fully justified by similar approaches pursued in the past centuries by classical methods of solving partial differential equations such as finite elements, finite differences, or spectral methods, where one would expand the unknown solution in terms of an appropriate set of basis functions. Approximating the unknown function $\eta$ by a deep neural network and using equation (\ref{eq:Structure}) allow us to obtain the following \emph{physics-informed neural network} (see figure \ref{fig:DeepVIV_1})
\begin{equation}
\begin{array}{l}
f_L := \rho \eta_{tt} + b \eta_t + k \eta.
\end{array}
\end{equation}
It is worth noting that the damping $b$ and the stiffness $k$ parameters turn into parameters of the resulting physics informed neural network $f_L$. We obtain the required derivatives to compute the residual network $f_L$ by applying the chain rule for differentiating compositions of functions using automatic differentiation \citep{baydin2015automatic}. Automatic differentiation is different from, and in several respects superior to, numerical or symbolic differentiation -- two commonly encountered techniques of computing derivatives. In its most basic description \citep{baydin2015automatic}, automatic differentiation relies on the fact that all numerical computations are ultimately compositions of a finite set of elementary operations for which derivatives are known. Combining the derivatives of the constituent operations through the chain rule gives the derivative of the overall composition. This allows accurate evaluation of derivatives at machine precision with ideal asymptotic efficiency and only a small constant factor of overhead. In particular, to compute the required derivatives we rely on Tensorflow \citep{abadi2016tensorflow}, which is a popular and relatively well documented open source software library for automatic differentiation and deep learning computations.\\

The shared parameters of the neural networks $\eta$ and $f_L$, in addition to the damping $b$ and the stiffness $k$ parameters, can be learned by minimizing the following sum of squared errors loss function
\begin{equation}\label{eq:loss_Structure}
\sum_{n=1}^N |\eta(t^n) - \eta^n|^2 + \sum_{n=1}^N |f_L(t^n) - f_L^n|^2.
\end{equation}
The first summation in this loss function corresponds to the training data on the displacement $\eta(t)$ while the second summation enforces the dynamics imposed by equation (\ref{eq:Structure}).

\subsection{Inferring Lift and Drag Forces from Scattered Velocity Measurements}

So far, we have been operating under the assumption that we have access to direct measurements of the lift force $f_L$. In the following, we are going to relax this assumption by recalling that the fluid motion is governed by the incompressible Navier-Stokes equations given explicitly by
\begin{equation}\label{eq:Fluid}
\begin{array}{l}
u_t + u u_x + v u_y = - p_x + \Rey^{-1}(u_{xx} + u_{yy}),\\
v_t + u v_x + v v_y = - p_y + \Rey^{-1}(v_{xx} + v_{yy}) - \eta_{tt},\\
u_x + v_y = 0.
\end{array}
\end{equation}
Here, $u(t,x,y)$ and $v(t,x,y)$ are the streamwise and crossflow components of the velocity field, respectively, while $p(t,x,y)$ denotes the pressure, and $\Rey$ is the Reynolds number based on the cylinder diameter and the free stream velocity. We consider the incompressible Navier-Stokes equations in the coordinate system attached to the cylinder, so that the cylinder appears stationary in time. This explains the appearance of the extra term $\eta_{tt}$ in the second momentum equation (\ref{eq:Fluid}).

\begin{figure}[!t]
\centering
\includegraphics[width=\textwidth]{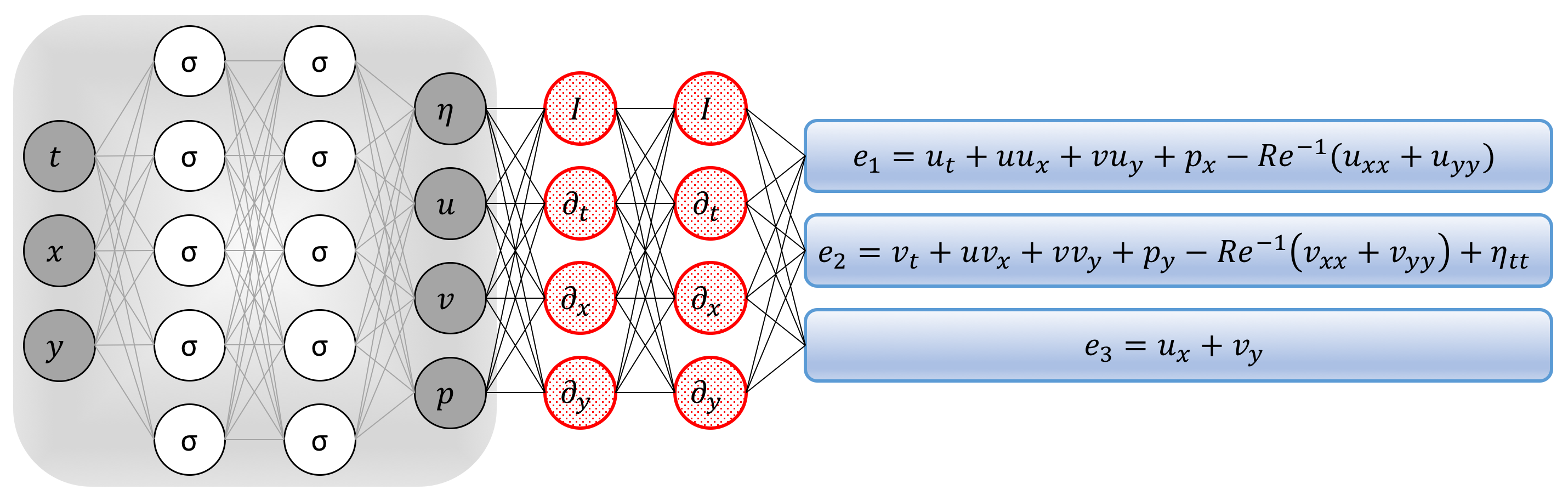}
\caption{\emph{Navier-Stokes informed neural networks:} A plain vanilla densely connected (physics uninformed) neural network, with 10 hidden layers and 32 neurons per hidden layer per output variable (i.e., $4 \times 32 = 128$ neurons per hidden layer), takes the input variables $t, x, y$ and outputs $\eta, u, v$, and $p$. As for the activation functions, we use $\sigma(x) = \sin(x)$. For illustration purposes only, the network depicted in this figure comprises of 2 hidden layers and 5 neurons per hidden layers. We employ automatic differentiation to obtain the required derivatives to compute the residual (physics informed) networks $e_1$, $e_2$, and $e_3$. If a term does not appear in the blue boxes (e.g., $u_{xy}$ or $u_{tt}$), its coefficient is assumed to be zero. It is worth emphasizing that unless the coefficient in front of a term is non-zero, that term is not going to appear in the actual ``compiled" computational graph and is not going to contribute to the computational cost of a feed forward evaluation of the resulting network. The total loss function is composed of the regression loss of the velocity fields $u, v$ and the displacement $\eta$ on the training data, and the loss imposed by the differential equations $e_1, e_2$, and $e_3$. Here, $I$ denotes the identity operator and the differential operators $\partial_t, \partial_x$, and $\partial_y$ are computed using automatic differentiation and can be thought of as ``activation operators". Moreover, the gradients of the loss function are back-propagated through the entire network to train the neural network parameters using the Adam optimizer.}\label{fig:DeepVIV_2}
\end{figure}

\begin{problem}[VIV-I]
Given scattered and potentially noisy measurements $\{t^n, x^n, y^n, u^n, v^n\}_{n=1}^N$ of the velocity field\footnote{Take for example the case of reconstructing a flow field from scattered measurements obtained from Particle Image Velocimetry (PIV) or Particle Tracking Velocimetry (PTV).} in addition to the data $\{t^n,\eta^n\}_{n=1}^{N}$ on the displacement and knowing the governing equations of the flow (\ref{eq:Fluid}), we are interested in reconstructing the entire velocity field as well as the pressure field in space-time. Such measurements are usually collected only in a small sub-domain, which may not be appropriate for classical CFD computations due to the presence of numerical artifacts. Typically, the data points are scattered in both space and time and are usually of the order of a few thousands or less in space.
\end{problem}

To solve the aforementioned problem, we proceed by approximating the latent functions $u(t,x,y)$, $v(t,x,y)$, $p(t,x,y)$, and $\eta(t)$ by a single neural network outputting four variables while taking as input $t, x$, and $y$. This prior assumption along with the incompressible Navier-Stokes equations (\ref{eq:Fluid}) result in the following \emph{Navier-Stokes informed neural networks} (see figure \ref{fig:DeepVIV_2})
\begin{equation}\label{eq:PINNs_NS}
\begin{array}{l}
e_1 := u_t + u u_x + v u_y + p_x - \Rey^{-1}(u_{xx} + u_{yy}),\\
e_2 := v_t + u v_x + v v_y + p_y - \Rey^{-1}(v_{xx} + v_{yy}) + \eta_{tt},\\
e_3 := u_x + v_y.
\end{array}
\end{equation}
We use automatic differentiation \citep{baydin2015automatic} to obtain the required derivatives to compute the residual networks $e_1$, $e_2$, and $e_3$. The shared parameters of the neural networks $u$, $v$, $p$, $\eta$, $e_1$, $e_2$, and $e_3$ can be learned by minimizing the sum of squared errors loss function
\begin{equation}
\arraycolsep=1.5pt\def\arraystretch{1.5}
\begin{array}{l}
\sum_{n=1}^N \left( \vert u(t^n,x^n,y^n)-u^n \vert^2 + \vert v(t^n,x^n,y^n)-v^n \vert^2 \right)\\
+ \sum_{n=1}^N \vert \eta(t^n)-\eta^n \vert^2 + \sum_{i=1}^3\sum_{n=1}^N \left( \vert e_i(t^n,x^n,y^n) \vert^2 \right).
\end{array}
\end{equation}
Here, the first two summations correspond to the training data on the fluid velocity and the structure displacement while the last summation enforces the dynamics imposed by equation (\ref{eq:Fluid}).\\

The fluid forces on the cylinder are functions of the pressure and the velocity gradients. Consequently, having trained the neural networks, we can use
\begin{eqnarray}
F_D = \oint \left[-p n_x + 2 \Rey^{-1} u_x n_x + \Rey^{-1} \left(u_y + v_x\right)n_y\right]ds,\label{eq:drag}\\
F_L = \oint \left[-p n_y + 2 \Rey^{-1} v_y n_y + \Rey^{-1} \left(u_y + v_x\right)n_x\right]ds,\label{eq:lift}
\end{eqnarray}
to obtain the lift and drag forces exerted by the fluid on the cylinder. Here, $(n_x,n_y)$ is the outward normal on the cylinder and $ds$ is the arc length on the surface of the cylinder. We use the trapezoidal rule to approximately compute these integrals, and we use equation (\ref{eq:lift}) to obtain the required data on the lift force. These data are then used to estimate the structural parameters $b$ and $k$ by minimizing the loss function (\ref{eq:loss_Structure}).

\subsection{Inferring Lift and Drag Forces from Flow Visualizations}

We now consider the second VIV learning problem by taking one step further and circumvent the need for having access to measurements of the velocity field by leveraging the following equation
\begin{equation}\label{eq:C}
c_t + u c_x + v c_y = \Pen^{-1} (c_{xx} + c_{yy}),
\end{equation}
governing the evolution of the concentration $c(t,x,y)$ of a passive scalar injected into the fluid flow dynamics described by the incompressible Navier-Stokes equations (\ref{eq:Fluid}). Here, $\Pen$ denotes the P\'{e}clet number, defined based on the cylinder diameter, the free-stream velocity and the diffusivity of the concentration species.

\begin{figure}[!t]
\centering
\includegraphics[width=\textwidth]{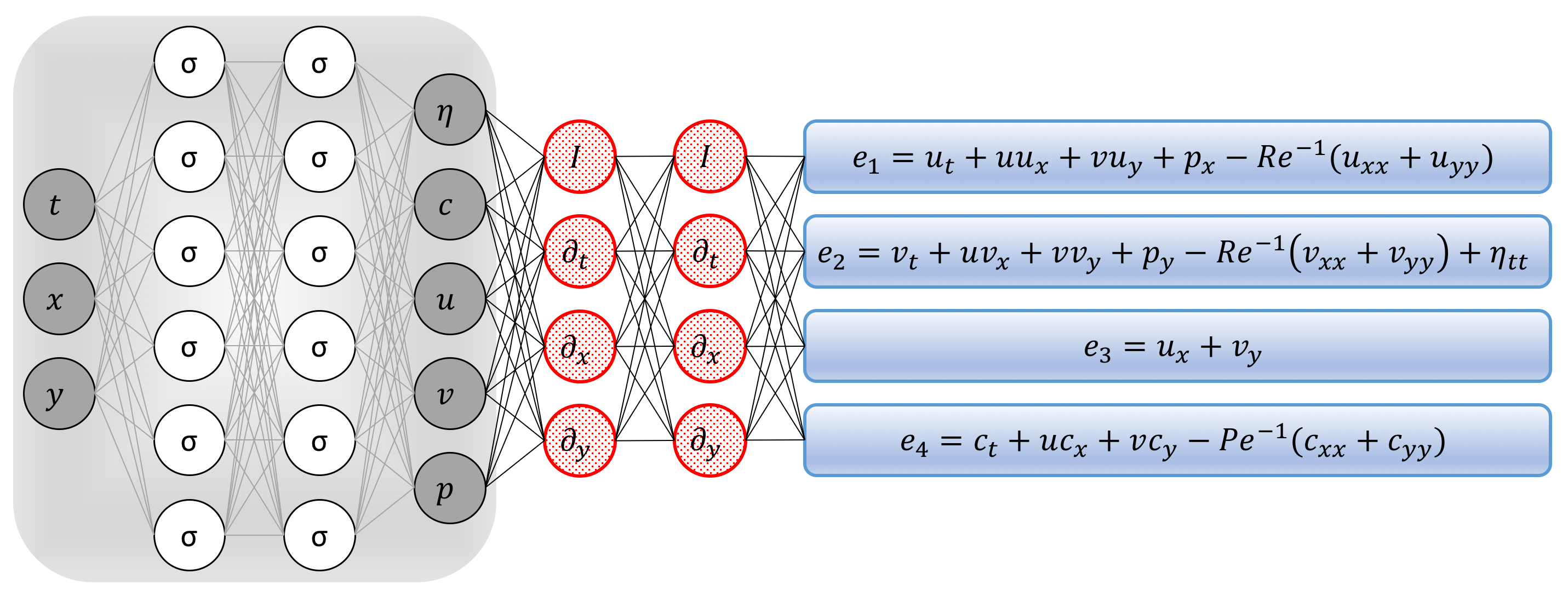}
\caption{\emph{Navier-Stokes informed neural networks:} A plain vanilla densely connected (physics uninformed) neural network, with 10 hidden layers and 32 neurons per hidden layer per output variable (i.e., $5 \times 32 = 160$ neurons per hidden layer), takes the input variables $t, x, y$ and outputs $\eta, c, u, v, w$, and $p$. As for the activation functions, we use $\sigma(x) = \sin(x)$. For illustration purposes only, the network depicted in this figure comprises of 2 hidden layers and 6 neurons per hidden layers. We employ automatic differentiation to obtain the required derivatives to compute the residual (physics informed) networks $e_1$, $e_2$, $e_3$, and $e_4$. If a term does not appear in the blue boxes (e.g., $u_{xy}$ or $u_{tt}$), its coefficient is assumed to be zero. It is worth emphasizing that unless the coefficient in front of a term is non-zero, that term is not going to appear in the actual ``compiled" computational graph and is not going to contribute to the computational cost of a feed forward evaluation of the resulting network. The total loss function is composed of the regression loss of the passive scalar $c$ and the displacement $\eta$ on the training data, and the loss imposed by the differential equations $e_1 - e_4$. Here, $I$ denotes the identity operator and the differential operators $\partial_t, \partial_x$, and $\partial_y$ are computed using automatic differentiation and can be thought of as ``activation operators". Moreover, the gradients of the loss function are back-propagated through the entire network to train the neural networks parameters using the Adam optimizer.}\label{fig:DeepVIV_3}
\end{figure}

\begin{problem}[VIV-II]
Given scattered and potentially noisy measurements $\{t^n,x^n,y^n,c^n\}_{n=1}^N$ of the concentration $c(t,x,y)$ of the passive scalar in space-time, we are interested in inferring the latent (hidden) quantities $u(t,x,y)$, $v(t,x,y)$, and $p(t,x,y)$ while leveraging the governing equations of the flow (\ref{eq:Fluid}) as well as equation (\ref{eq:C}) describing the evolution of the passive scalar. Typically, the data points are of the order of a few thousands or less in space. Moreover, equations (\ref{eq:drag}) and (\ref{eq:lift}) enable us to consequently compute the drag and lift forces, respectively, as functions of the inferred pressure and velocity gradients. Unlike the first VIV problem, here we assume that we do not have access to direct observations of the velocity field.
\end{problem}

To solve the second VIV problem, in addition to approximating $u(t,x,y)$, $v(t,x,y)$, $p(t,x,y)$, and $\eta(t)$ by deep neural networks as before, we represent $c(t,x,y)$ by yet another output of the network taking $t, x,$ and $y$ as inputs. This prior assumption along with equation (\ref{eq:C}) results in the following additional component of the \emph{Navier-Stokes informed neural network} (see figure \ref{fig:DeepVIV_3})
\begin{equation}
\begin{array}{l}
e_4 := c_t + u c_x + v c_y - \Pen^{-1}(c_{xx} + c_{yy}).
\end{array}
\end{equation}

The residual networks $e_1$, $e_2$, and $e_3$ are defined as before according to equation (\ref{eq:PINNs_NS}). We use automatic differentiation \citep{baydin2015automatic} to obtain the required derivatives to compute the additional residual network $e_4$. The shared parameters of the neural networks $c$, $u$, $v$, $p$, $\eta$, $e_1$, $e_2$, $e_3$, and $e_4$ can be learned by minimizing the sum of squared errors loss function
\begin{equation}
\arraycolsep=1.5pt\def\arraystretch{1.5}
\begin{array}{l}
\sum_{n=1}^N \left(\vert c(t^n,x^n,y^n)-c^n \vert^2 + \vert \eta(t^n)-\eta^n \vert^2\right)\\
+ \sum_{m=1}^M \left( \vert u(t^m,x^m,y^m)-u^m \vert^2 + \vert v(t^m,x^m,y^m)-v^m \vert^2 \right)\\
+ \sum_{i=1}^4\sum_{n=1}^N \left( \vert e_i(t^n,x^n,y^n) \vert^2 \right).
\end{array}
\end{equation}
Here, the first summation corresponds to the training data on the concentration of the passive scalar and the structure's displacement, the second summation corresponds to the Dirichlet boundary data on the velocity field, and the last summation enforces the dynamics imposed by equations (\ref{eq:Fluid}) and (\ref{eq:C}). Upon training, we use equation (\ref{eq:lift}) to obtain the required data on the lift force. Such data are then used to estimate the structural parameters $b$ and $k$ by minimizing the loss function (\ref{eq:loss_Structure}).

\section[Results]{Results\protect\footnote{All data and codes used in this manuscript will be publicly available on GitHub at https://github.com/maziarraissi/DeepVIV.}}\label{sec:Results}

To generate a high-resolution dataset for the VIV problem we have performed direct numerical simulations (DNS) employing the high-order spectral-element method \citep{GK_book}, together with the coordinate transformation method to take account of the boundary deformation \citep{NEWMAN_viv}. The computational domain is $[-6.5\,D,23.5\,D] \times [-10\,D,10 \, D]$, consisting of 1,872 quadrilateral elements. The cylinder center was placed at $(0, 0)$. On the inflow, located at $x/D=-6.5$, we prescribe $(u=U_{\infty},v=0)$. On the outflow, where $x/D=23.5$, zero-pressure boundary condition $(p=0)$ is imposed. On both top and bottom boundaries where $y/D=\pm 10$, a periodic boundary condition is used. The Reynolds number is $\Rey=100$, $\rho=2$, $b=0.084$ and $k=2.2020$. For the case with dye, we assumed the P\'{e}clet number $\Pen=90$. First, the simulation is carried out until $t=1000 \, \frac{D}{U_{\infty}}$ when the system is in steady periodic state. Then, an additional simulation for $\Delta t=14 \,\frac{D}{U_{\infty}}$ is performed to collect the data that are saved in 280 field snapshots. The time interval between two consecutive snapshots is $\Delta t= 0.05 \frac{D}{U_{\infty}}$. Note here $D=1$ is the diameter of the cylinder and $U_{\infty}=1$ is the inflow velocity. We use the DNS results to compute the lift and drag forces exerted by the fluid on the cylinder. All data and codes used in this manuscript will be publicly available on GitHub at \url{https://github.com/maziarraissi/DeepVIV}.

\subsection{A Pedagogical Example}

\begin{figure}[!t]
\centering
\includegraphics[width=\textwidth]{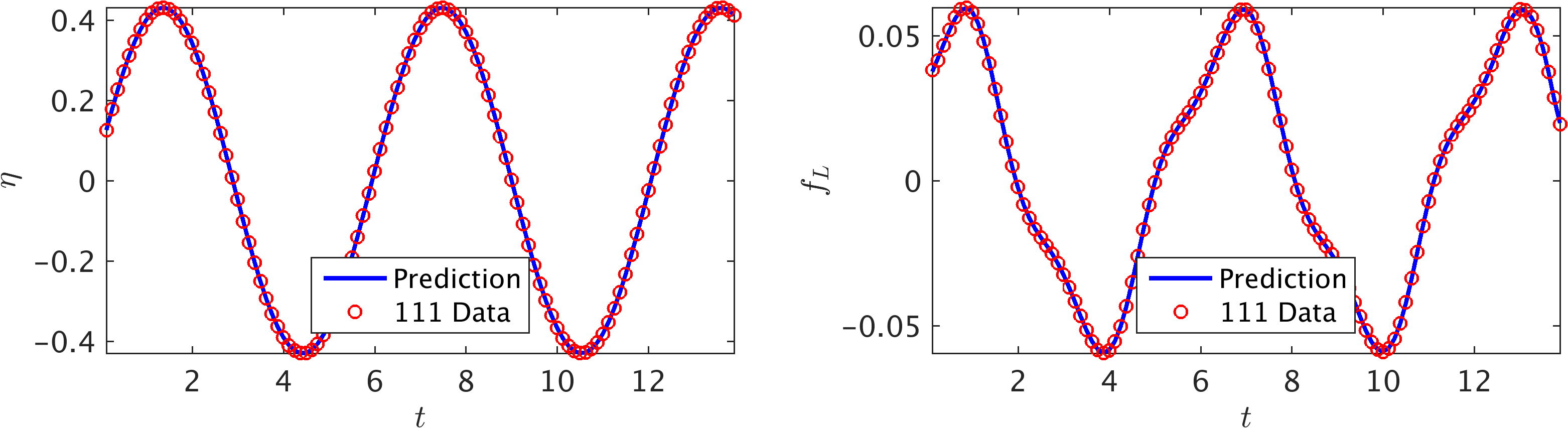}
\caption{\emph{Vortex Induced Vibrations}: Observations of the displacement $\eta$ are plotted in the left panel while the data on the lift force $f_L$ are depicted in the right panel. These observations are shown by the red circles. Predictions of the trained neural networks $\eta$ and $f_L$ are depicted by blue solid lines.}\label{fig:displacement_lift}
\end{figure}

To illustrate the effectiveness of our approach, let us start with the two time series depicted in figure \ref{fig:displacement_lift} consisting of $N=111$ observations of the displacement and the lift force. These data correspond to damping and stiffness parameters with exact values $b=0.084$ and $k=2.2020$, respectively. Here, the cylinder is assumed to have a mass of $\rho = 2.0$. This data-set is then used to train a 10-layer deep neural network with 32 neurons per hidden layers (see figure \ref{fig:DeepVIV_1}) by minimizing the sum of squared errors loss function (\ref{eq:loss_Structure}) using the Adam optimizer \citep{kingma2014adam}. Upon training, the network is used to predict the entire solution functions $\eta(t)$ and $f_L(t)$, as well as the unknown structural parameters $b$ and $k$. In addition to almost perfect reconstructions of the two time series for displacement and lift force, the proposed framework is capable of identifying the correct values for the structural parameters $b$ and $k$ with remarkable accuracy. The learned values for the damping and stiffness parameters are $b = 0.08438281$ and $k = 2.2015007$. This corresponds to around $0.45\%$ and $0.02\%$ relative errors in the estimated values for $b$ and $k$, respectively.\\

As for the activation functions, we use $\sin(x)$. In general, the choice of a neural network's architecture (e.g., number of layers/neurons and form of activation functions) is crucial and in many cases still remains an art that relies on one's ability to balance the trade off between \emph{expressivity} and \emph{trainability} of the neural network \citep{raghu2016expressive}. Our empirical findings so far indicate that deeper and wider networks are usually more expressive (i.e., they can capture a larger class of functions) but are often more costly to train (i.e., a feed-forward evaluation of the neural network takes more time and the optimizer requires more iterations to converge). Moreover, the sinusoid (i.e., $\sin(x)$) activation function seems to be numerically more stable than $\tanh(x)$, at least while computing the residual neural networks $f_L$, and $e_i$, $i=1,\ldots,4$ (see figures \ref{fig:DeepVIV_1}, \ref{fig:DeepVIV_2}, and \ref{fig:DeepVIV_3}). In this work, we have tried to choose the neural networks' architectures in a consistent fashion throughout the manuscript by setting the number of layers to 10 and the number of neurons to 32. Consequently, there might exist other architectures that improve some of the results reported in the current work.

\subsection{Inferring Lift and Drag Forces from Scattered Velocity Measurements}

\begin{figure}[!t]
\centering
\includegraphics[width=\textwidth]{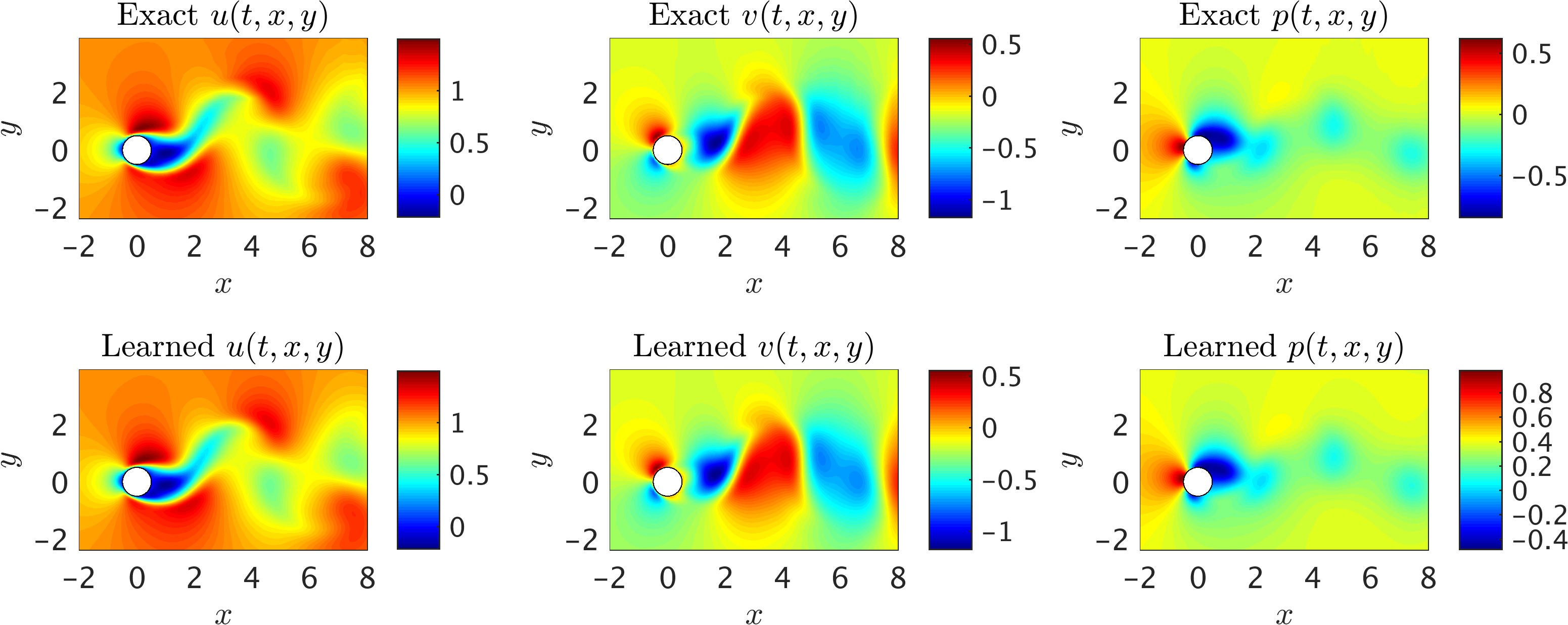}
\caption{\emph{VIV-I (Velocity Measurements)}: A representative snapshot of the data on the velocity field is depicted in the top left and top middle panels of this figure. The algorithm is capable of accurately (of the order of $10^{-3}$) reconstructing the velocity field and more importantly the pressure without having access to even a single observation on the pressure itself.}\label{fig:VIV_Case2_data_on_velocities_results}
\end{figure}

\begin{figure}[!t]
\includegraphics[width=\textwidth]{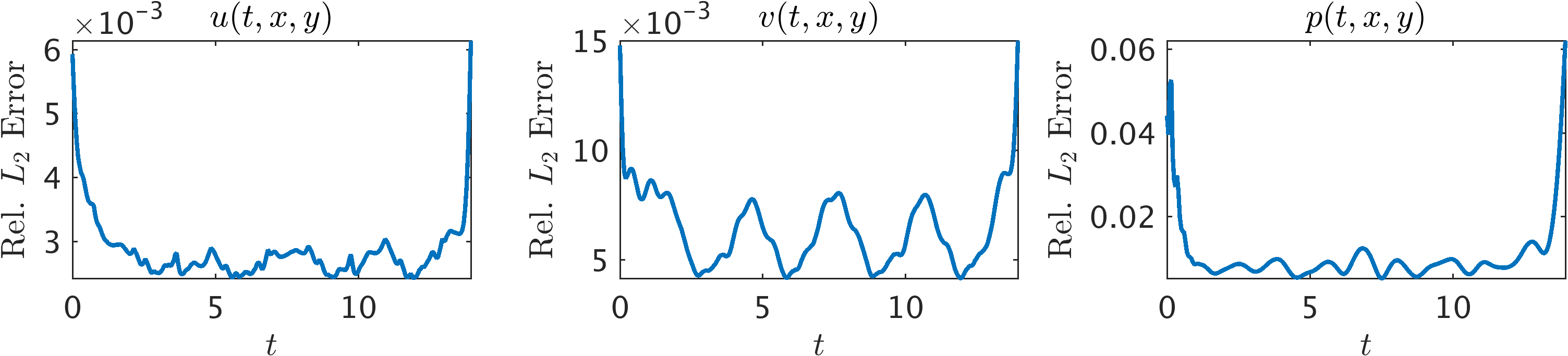}
\caption{\emph{VIV-I (Velocity Measurements)}: Relative $L_2$ errors between predictions of the model and the corresponding exact velocity and pressure fields. 4 million data points corresponding 280 time snapshots, scattered in space and time, are used both to regress the velocity field and enforce the corresponding partial differential equations. Lack of training data on $u$ and $v$ for $t < 0$ and $t > 14$ leads to weaker neural network predictions for the initial and final time instants.}\label{fig:VIV_data_on_velocities_errors}
\end{figure}

Let us now consider the case where we do not have access to direct measurements of the lift force $f_L$. In this case, we can use measurements of the velocity field, obtained for instance via Particle Image Velocimetry (PIV) or Particle Tracking Velocimetry (PTV), to reconstruct the velocity field, the pressure, and consequently the drag and lift forces. A representative snapshot of the data on the velocity field is depicted in the top left and top middle panels of figure \ref{fig:VIV_Case2_data_on_velocities_results}. The neural network architectures used here consist of 10 layers with 32 neurons in each hidden layer. A summary of our results is presented in figure \ref{fig:VIV_Case2_data_on_velocities_results}. The proposed framework is capable of accurately (of the order of $10^{-3}$) reconstructing the velocity field; however, a more intriguing result stems from the network's ability to provide an accurate prediction of the entire pressure field $p(t,x,y)$ in the absence of any training data on the pressure itself (see figure \ref{fig:VIV_data_on_velocities_errors}). A visual comparison against the exact pressure is presented in figure \ref{fig:VIV_Case2_data_on_velocities_results} for a representative snapshot of the pressure. It is worth noticing that the difference in magnitude between the exact and the predicted pressure is justified by the very nature of incompressible Navier-Stokes equations, since the pressure field is only identifiable up to a constant. This result of inferring a continuous quantity of interest from auxiliary measurements by leveraging the underlying physics is a great example of the enhanced  capabilities that our approach has to offer, and highlights its potential in solving high-dimensional data assimilation and inverse problems.\\

\begin{figure}[!t]
\centering
\includegraphics[width=\textwidth]{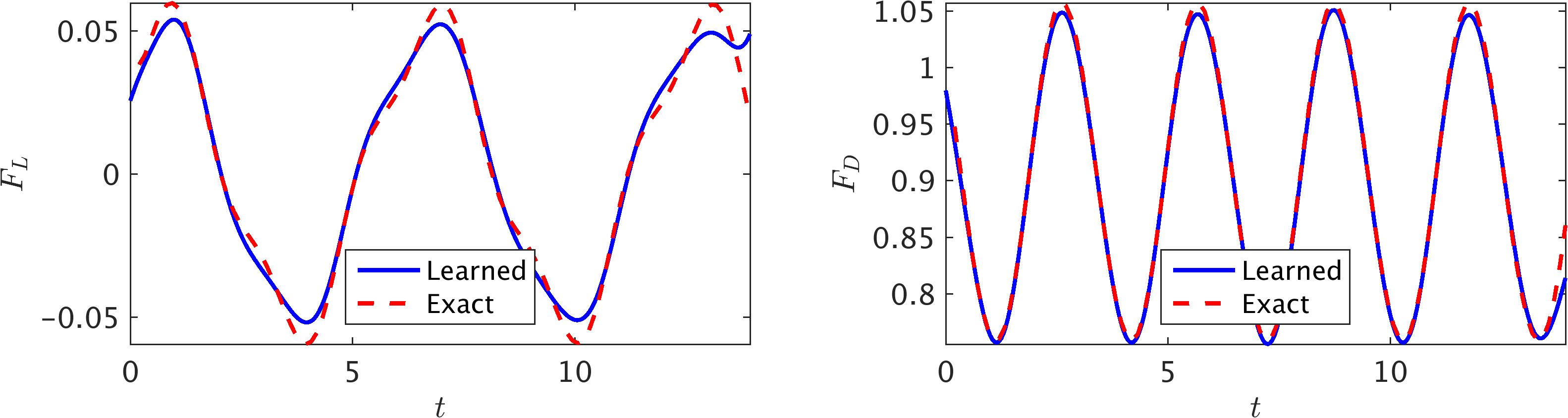}
\caption{\emph{VIV-I (Velocity Measurements)}: In this figure, the resulting lift (left) and drag (right) forces are compared to the exact ones.}\label{fig:VIV_data_on_velocities_lift_drag_results}
\end{figure}

The trained neural networks representing the velocity field and the pressure can be used to compute the drag and lift forces by employing equations (\ref{eq:drag}) and (\ref{eq:lift}), respectively. The resulting drag and lift forces are compared to the exact ones in figure \ref{fig:VIV_data_on_velocities_lift_drag_results}. In the following, we are going to use the computed lift force to generate the required training data on $f_L$ and estimate the structural parameters $b$ and $k$ by minimizing the loss function (\ref{eq:loss_Structure}). Upon training, the proposed framework is capable of identifying the correct values for the structural parameters $b$ and $k$ with remarkable accuracy. The learned values for the damping and stiffness parameters are $b = 0.0844064$ and $k = 2.1938791$. This corresponds to around $0.48\%$ and $0.37\%$ relative errors in the estimated values for $b$ and $k$, respectively.

\subsection{Inferring Lift and Drag Forces from Flow Visualizations}

\begin{figure}[!t]
\centering
\includegraphics[width=\textwidth]{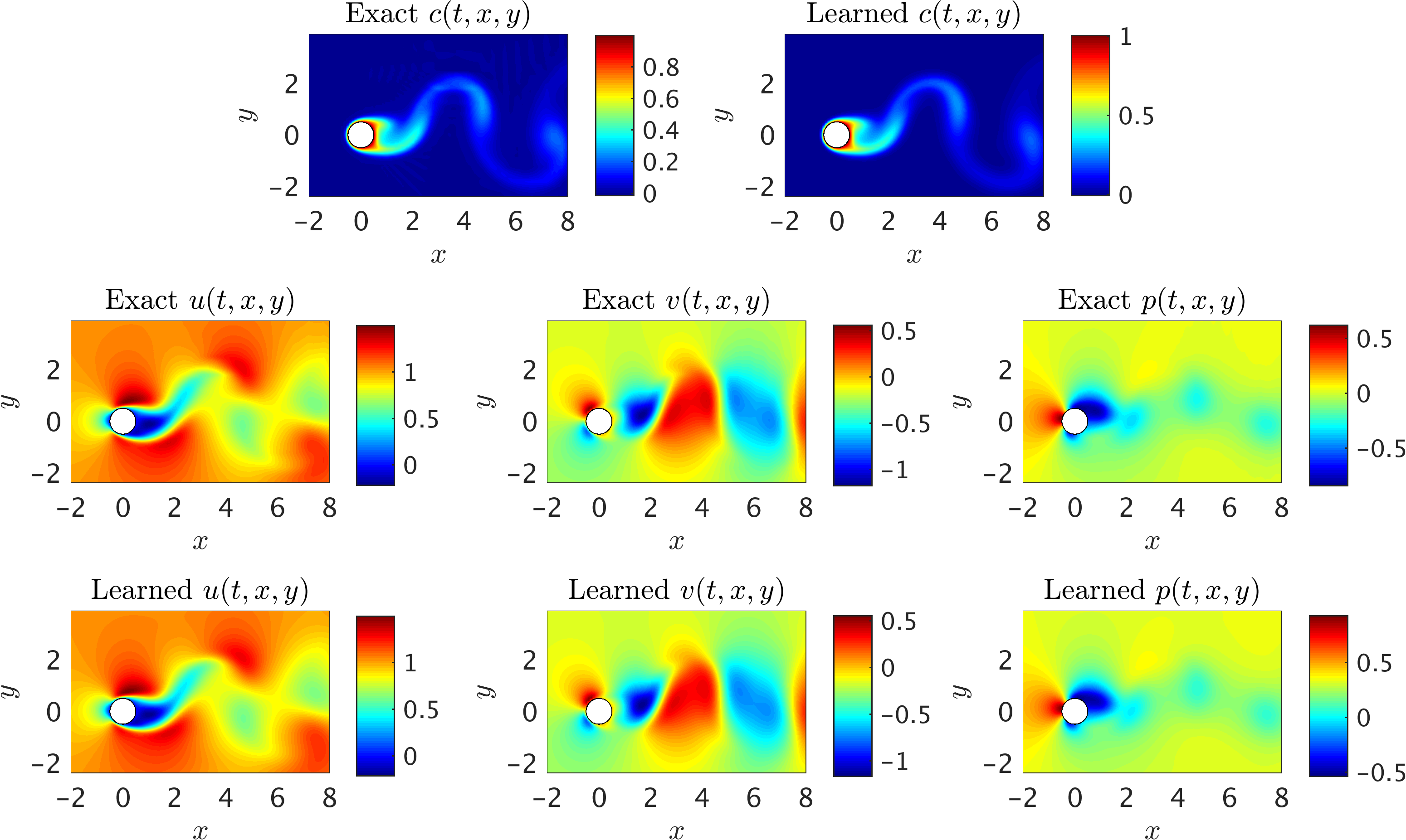}
\caption{\emph{VIV-II (Flow Visualizations Data)}: A representative snapshot of the data on the concentration of the passive scalar is depicted in the top left panel of this figure. The algorithm is capable of accurately (of the order of $10^{-3}$) reconstructing the concentration of the passive scalar and more importantly the velocity field as well as the pressure without having access to enough observations of these quantities themselves.}\label{fig:VIV_case2_concentration_results}
\end{figure}

\begin{figure}[!t]
\includegraphics[width=\textwidth]{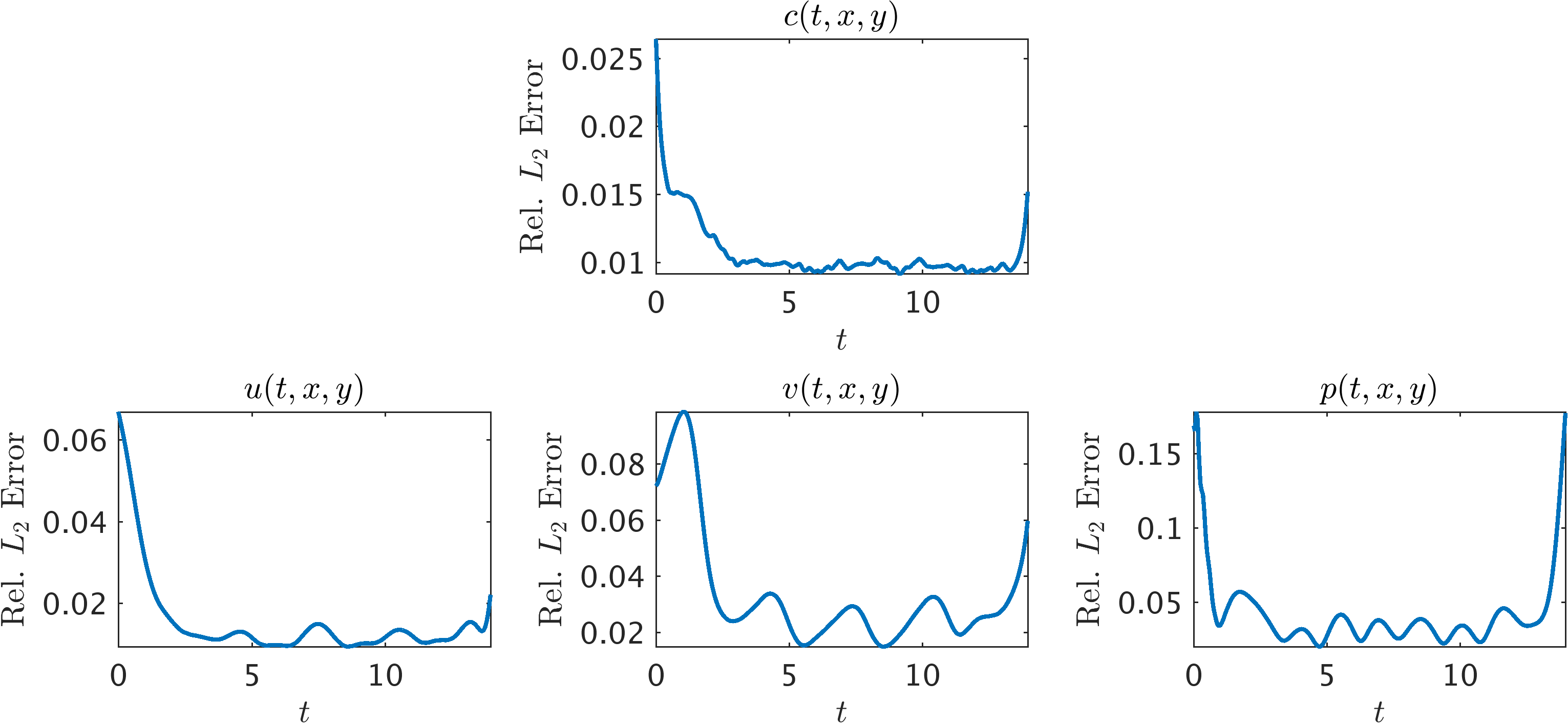}
\caption{\emph{VIV-II (Flow Visualizations Data)}: Relative $L_2$ errors between predictions of the model and the corresponding exact velocity and pressure fields. 4 million data points corresponding to 280 time snapshots, scattered in space and time, are used both to regress the concentration field and enforce the corresponding partial differential equations. Lack of training data on $c$ for $t < 0$ and $t > 14$ leads to weaker neural network predictions for the initial and final time instants.}\label{fig:VIV_data_on_concentration_errors}
\end{figure}

Let us continue with the case where we do not have access to direct observations of the lift force $f_L$. This time rather than using data on the velocity field, we use measurements of the concentration of a passive scalar (e.g., dye or smoke) injected into the system. In the following, we are going to employ such data to reconstruct the velocity field, the pressure, and consequently the drag and lift forces. A representative snapshot of the data on the concentration of the passive scalar is depicted in the top left panel of figure \ref{fig:VIV_case2_concentration_results}. The neural networks' architectures used here consist of 10 layers with 32 neurons per each hidden layer. A summary of our results is presented in figure \ref{fig:VIV_case2_concentration_results}. The proposed framework is capable of accurately (of the order of $10^{-3}$) reconstructing the concentration. However, a truly intriguing result stems from the network's ability to provide accurate predictions of the entire velocity vector field as well as the pressure, in the absence of sufficient training data on the pressure and the velocity field themselves (see figure \ref{fig:VIV_data_on_concentration_errors}). A visual comparison against the exact quantities is presented in figure \ref{fig:VIV_case2_concentration_results} for a representative snapshot of the velocity field and the pressure. This result of inferring multiple hidden quantities of interest from auxiliary measurements by leveraging the underlying physics is a great example of the enhanced capabilities that \emph{physics-informed deep learning} has to offer, and highlights its potential in solving high-dimensional data-assimilation and inverse problems (see e.g., \cite{raissi2018hiddenfluid}).\\

\begin{figure}
\centering
\includegraphics[width=\textwidth]{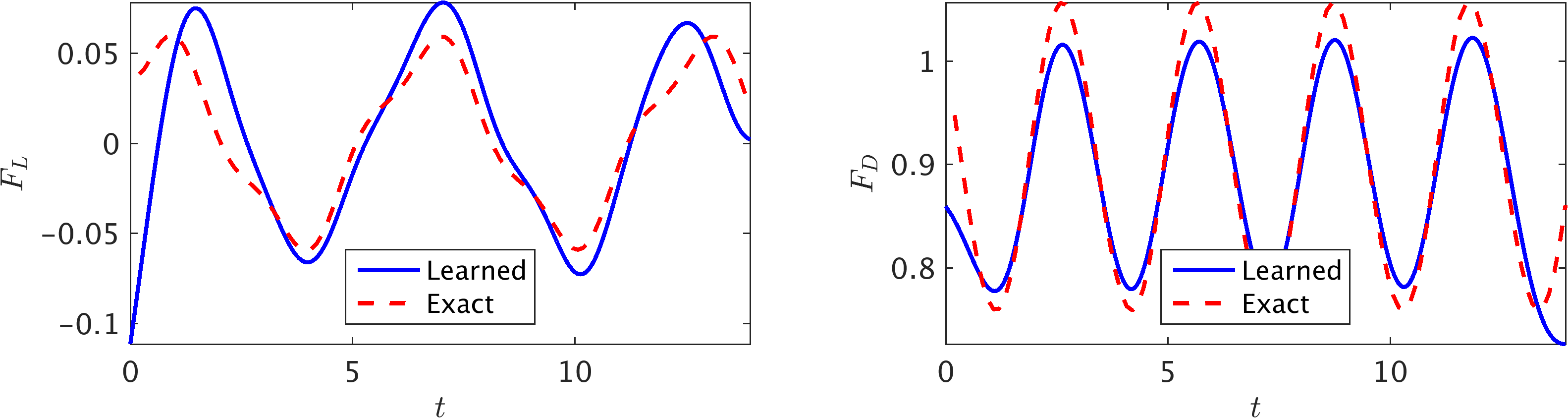}
\caption{\emph{Vortex Induced Vibrations (Flow Visualizations Data)}: In this figure, the resulting lift (left) and drag (right) forces are compared to the exact ones.}\label{fig:VIV_data_on_concentration_lift_drag_results}
\end{figure}

Following the same procedure as in the previous example, the trained neural networks representing the velocity field and the pressure can be used to compute the drag and lift forces by employing equations (\ref{eq:drag}) and (\ref{eq:lift}), respectively. The resulting drag and lift forces are compared to the exact ones in figure \ref{fig:VIV_data_on_concentration_lift_drag_results}. In the following, we are going to use the computed lift force to generate the required training data on $f_L$ and estimate the structural parameters $b$ and $k$ by minimizing the loss function (\ref{eq:loss_Structure}). Upon training, the proposed framework is capable of identifying the correct values for the structural parameters $b$ and $k$ with surprising accuracy. The learned values for the damping and stiffness parameters are $b = 0.08600664$ and $k = 2.2395933$. This corresponds to around $2.39\%$ and $1.71\%$ relative errors in the estimated values for $b$ and $k$, respectively.

\section{Discussion and Concluding Remarks}

We have considered the classical coupled problem of a freely vibrating cylinder due to lift forces and demonstrated how physics informed deep learning can be used to infer quantities of interest from scattered data in space-time. In the {\em first VIV learning problem}, we inferred the pressure field and structural parameters, and hence the lift and drag on the vibrating cylinder using velocity and displacement data in time-space. In the {\em second VIV learning problem}, we inferred the velocity and pressure fields as well as the structural parameters given data on a passive scalar in space-time. The framework we propose here represents a {\em paradigm shift} in fluid mechanics simulation as it uses the governing equations as regularization mechanisms in the loss function of the corresponding minimization problem. It is particularly effective for multi-physics problems as the coupling between fields can be readily accomplished by sharing parameters among the  multiple neural networks -- here a neural network outputting 4 variables for the first problem and 5 variables for the second one -- and for more general coupled problems by also including coupled terms in the loss function. There are many questions that this new type of modeling raises, both theoretical and practical, e.g. efficiency, solution uniqueness, accuracy, etc. We have considered such questions here in the present context as well as in our previous work in the context of physics-informed learning machines but admittedly at the present time it is not possible to rigorously answer such questions. We hope, however, that our present work will ignite interest in physics-informed deep learning that can be used effectively for many different fields of multi-physics fluid mechanics.\\

The results presented in the current work are among those cases where a pure machine learning algorithm or a mere scientific computing approach simply cannot reproduce. A pure machine learning strategy has no sense of the physics of the problem to begin with, and a mere scientific computing approach relies heavily on careful specification of the geometry as well as initial and boundary conditions. Assuming the geometry to be known, to arrive at similar results as the ones presented in the current work, one needs to solve significantly more expensive optimization problems, using conventional computational methods (e.g., finite differences, finite elements, finite volumes, spectral methods, and etc.). The corresponding optimization problems involve some form of ``parametrized" initial and boundary conditions, appropriate loss functions, and multiple runs of the conventional computational solvers. In this setting, one could easily end up with very high-dimensional optimization problems that require either backpropagating through the computational solvers \cite{chen2018neural} or ``Bayesian" optimization techniques \cite{shahriari2016taking} for the surrogate models (e.g., Gaussian processes). If the geometry is further assumed to be unknown (as is the case in this work), then its parametrization requires grid regeneration, which makes the approach almost impractical.\\

However, it must be mentioned that we are avoiding the regimes where the Navier-Stokes equations become chaotic and turbulent (e.g., as the Reynolds number increases). In fact, it should not be difficult for a plain vanilla neural network to approximate the types of complicated functions that naturally appear in turbulence. However, as we compute the derivatives required in the computation of the physics informed neural networks (see figures \ref{fig:DeepVIV_1}, \ref{fig:DeepVIV_2}, and \ref{fig:DeepVIV_3}), minimizing the loss functions might become a challenge \cite{raissi2018deep}, where the optimizer may fail to converge to the right values for the parameters of the neural networks. It might be the case that the resulting optimization problem inherits the complicated nature of the turbulent Navier-Stokes equations. Hence, inference of turbulent velocity and pressure fields should be considered in future extensions of this line of research. Moreover, in this work we have been operating under the assumption of Newtonian and incompressible fluid flow governed by the Navier-Stokes equations. However, the proposed algorithm can also be used when the underlying physics is non-Newtonian, compressible, or partially known. This, in fact, is one of the advantages of our algorithm in which other unknown parameters such as the Reynolds and P\'eclet numbers can be inferred in addition to the velocity and pressure fields.

\section*{Acknowledgements}
This work received support by the DARPA EQUiPS grant N66001-15-2-4055 and the AFOSR grant FA9550-17-1-0013. All data and codes used in this manuscript will be publicly available on GitHub at \url{https://github.com/maziarraiss/DeepVIV}.





\bibliographystyle{model1-num-names}
\bibliography{sample.bib}







\end{document}